\documentclass[runningheads]{llncs}
\usepackage{algorithm,color,hyperref,url}

\begin{document}

\title{Rpair: Rescaling RePair with Rsync \thanks{Partially funded with Basal Funds FB0001, Conicyt, Chile.}}

\author{Travis Gagie\inst{1,2} \and
Tomohiro I\inst{3} \and
Giovanni Manzini\inst{4} \and
Gonzalo Navarro\inst{1,5} \and\\
Hiroshi Sakamoto\inst{3} \and
Yoshimasa Takabatake\inst{3}}
\authorrunning{Gagie, I, Manzini, Navarro, Sakamoto and Takabatake}
\institute{CeBiB --- Center for Biotechnology and Bioengineering, Chile \and
Faculty of Computer Science, Dalhousie University, Canada \and
Department of Artificial Intelligence,\\
Kyushu Institute of Technology, Fukuoka, Japan \and
Department of Science and Technological Innovation,\\
University of Eastern Piedmont, Alessandria, Italy \and
Department of Computer Science,
University of Chile, Santiago, Chile}
\maketitle

\begin{abstract}
Data compression is a powerful tool for managing massive but repetitive datasets, especially schemes such as grammar-based compression that support computation over the data without decompressing it.  In the best case such a scheme takes a dataset so big that it must be stored on disk and shrinks it enough that it can be stored and processed in internal memory.  Even then, however, the scheme is essentially useless unless it can be built on the original dataset reasonably quickly while keeping the dataset on disk.  In this paper we show how we can preprocess such datasets with context-triggered piecewise hashing such that afterwards we can apply RePair and other grammar-based compressors more easily.  We first give our algorithm, then show how a variant of it can be used to approximate the LZ77 parse, then leverage that to prove theoretical bounds on compression, and finally give experimental evidence that our approach is competitive in practice.
\end{abstract}

\section{Introduction}
\label{sec:introduction}

Dictionary compression has proved to be an effective tool to exploit the repetitiveness that most of the fastest-growing datasets feature \cite{Nav12}. Lempel-Ziv (LZ77 for short) \cite{LZ76,ZL77} stands out as the most popular and effective compression method for repetitive texts. Further, it can be run in linear time and even in external memory \cite{KKP14}. LZ77 has the important drawback, however, that accessing random positions of the compressed text requires, essentially, to decompress it from the beginning. Therefore, it is not suitable to be used as a {\em compressed data structure} that represents the text in little space while simulating direct access to it.
Grammar compression \cite{KY00} is an alternative that offers better guarantees in this sense. The aim is to build a small context-free grammar (or Straight-Line Program, SLP) that generates (only) the text. The smallest SLP generating a text is always larger than its LZ77 parse, but only by a logarithmic factor that is rarely reached in practice. With an SLP we can access any text substring with only an additive logarithmic time penalty \cite{BPT15,BLRSRW15}, which has led to the development of various self-indexes building on SLPs \cite{PhBiCPM17,CE18,CN10,CN12,gagie2014lz77,NIIBT15}. In addition, many other richer queries on sequences have been supported by associating summary information the nonterminals of the SLP \cite{ACN13,BannaiGI18,BLRSRW15,BGBNP19,CM13,CFMPN16}.

Although finding the smallest SLP for a text is NP-complete \cite{CLLPPSS05,Ryt03}, there are several grammar construction algorithms that guarantee at most a logarithmic blowup on the LZ77 parse \cite{CLLPPSS05,Jez15,Jez16,Ryt03,Sak05}. In practice, however, they are sharply outperformed by RePair \cite{LM00}, a heuristic that runs in linear time and obtains grammars of size very close to that of the LZ77 parse in most cases. This has made RePair the compressor of choice to build grammar-based compressed data structures \cite{ACN13,BGBNP19,CFMPN16,CM13}. A serious problem with RePair, however, is that, despite running in linear time and space, in practice the constant is high and it can be built only on inputs that are about one tenth of the available memory. This significantly hampers its applicability on large datasets.

In this paper we introduce a scalable SLP compression algorithm that obtains space very close to that of RePair and can be applied on very large inputs. We prove a constant-approximation factor with respect to any SLP construction algorithm to which our technique is applied. Our experimental results show that we can compress a very repetitive 50GB text in less than an hour, using less than 650MB of RAM and obtaining very competitive compression ratios.

\section{Preliminaries}
\label{sec:preliminaries}

For the sake of brevity, we assume the reader is familiar with SLPs, LZ77, and the links between the two.  To prove theoretical bounds for our approach, we consider a variant of LZ77 in which  if $S [i..j]$ is  a  phrase  then  either $i = j$ and $S [i]$ is the first occurrence of a distinct character, or $S [i..j]$ occurs in $S [1..j - 1]$ and $S [i..j + 1]$ does not occur in $S [1..j]$.  We refer to this variant as LZSS due to its similarity to Storer and Szymanski's version of LZ77~\cite{LZSS1982StorerS}, even though they allow substrings to be stored as raw text and we do not.

The best-known algorithm for building SLPs is probably RePair~\cite{LM00}, for which there are many implementations (see~\cite{FuruyaTNIBK2019} and references therein).  It works by repeatedly finding the most common pair of symbols and replacing them with a new non-terminal.  Although it is not known to have a good worst-case approximation ratio with respect to the size of LZ77 parsing,  in practice it outperforms other constructions.  RePair uses linear time and space but the coefficient in the space bound is quite large and so the standard implementations are practical only on small inputs. A more recent and more space economical alternative to RePair is {SOLCA}~\cite{TakabatakeIS2017} that we will consider in Section~\ref{sec:experiments}.

Context-triggered piecewise hashing (CTPH) is a technique for parsing strings into blocks such that long repeated substrings are parsed the same way (except possibly at the beginning or end of the substrings).  The name CTPH seems to be due to to Kornblum~\cite{Kornblum2006} but the ideas go back to Tridgell's Rsync~\cite{Tridgell1999} and Spamsum (\url{https://www.samba.org/ftp/unpacked/junkcode/spamsum/README}): \\
``The core of the spamsum algorithm is a rolling hash similar to the rolling hash used in `rsync'. The rolling hash is used to produce a series of 'reset points' in the plaintext that depend only on the immediate context (with a default context width of seven characters) and not on the earlier or later parts of the plaintext.''

Specifically, in this paper we choose a rolling hash function and a threshold $p$, run a sliding window of fixed size $w$ over $S$ and end the current block whenever the window contains a triggering substring, which is a substring of length $w$ whose hash is congruent to 0 modulo $p$.  When we end a block, we shift the window ahead $w$ characters so all the blocks are disjoint and form a parse, 
which we call the {\em Rsync parse}. We call the set of distinct blocks the {\em Rsync dictionary}: if the input text contains many repetitions, we expect the dictionary to be much smaller than the text.

\section{Algorithms}
\label{sec:algorithms}

Given a string $S$, we can use Rsync parsing to help build an SLP for $S$ with Algorithm~\ref{alg:rpair} (``Rpair'').  The final SLP can be viewed as first generating the parse, then replacing each block ID in the parse by the sublist of non-terminals that generate each block, and finally replacing the sublists by the blocks themselves.

\begin{algorithm}[tb]
\caption{Rpair: use Rsync parsing to build an SLP for a given string $S$}
\label{alg:rpair}
\begin{enumerate}
\item build an Rsync dictionary and parse for $S$;
\item generate SLPs for the distinct blocks as follows:
\begin{enumerate}
\item append a unique separator character to each block in the dictionary and then concatenate the blocks (in the order of their first appearances in $S$) into a string $D$;
\item build an SLP for $D$;
\label{item:SLP_D}
\item delete from the SLP any non-terminal that occurs only once in the parse tree (and any rule including it);
\label{item:deletion}
\item delete from the SLP the separator characters (and any rules including them);
\item list the non-terminals at the roots of the maximal remaining subtrees of the parse tree;
\item divide the list into sublists such that the concatenation of the expansions of the non-terminals in the $i$th sublist is the $i$ block in $D$;
\item create a set of rules generating the $i$th sublist from a new non-terminal $X_i$;
\label{item:Xi}
\end{enumerate}
\item build an SLP for the parse $P$;
\label{item:SLP_P}
\item replace by $X_i$ each occurrence in $P$ of the terminal for the $i$th block in $D$;
\label{item:replace}
\item combine the SLP for $P$ with the SLPs for the blocks.
\label{item:combine}
\end{enumerate}
\end{algorithm}

Since each separator character appears only once in $D$ and its parse tree, any non-terminal whose expansion includes a separator character also appears only once and is deleted.  Since the parse tree of an SLP is binary and each non-terminal we delete appears only once, the number of distinct non-terminals we delete is at least the length of the list of non-terminals at the roots of the maximal remaining subtrees of the parse tree, minus one.  Therefore, creating rules to generate the sublists does not cause the number of distinct non-terminals to grow to more than the number in the original SLP for $D$, plus one.

Algorithm~\ref{alg:rpair} works with any algorithm for building SLPs for $D$ and $P$.  In Section~\ref{sec:analysis} we show that, if we choose an algorithm that builds SLPs for $D$ and $P$ at most an $\alpha$-factor larger than their LZ77 parses, then we obtain an SLP an $O (\alpha)$-factor larger than the LZ77 parse of $S$.  In the process we will refer to Algorithm~\ref{alg:rparse} (``Rparse''), which produces an LZSS-like parse of $S$ but is intended only to simplify our analysis of Algorithm~\ref{alg:rpair} (not to compete with cutting-edge LZ-based compressors).  By ``LZSS-like'' we mean a parse in which each phrase is either a single character that has not occurred before, or a copy of an earlier substring.  We note in passing that, if the parse in Step~\ref{item:SLP_P} is still to big to for a normal construction, then we can apply Algorithm~\ref{alg:rpair} to it.  We will show in the full version of this paper that, if we recurse only a constant number of times, then we worsen our compression bounds by only a constant factor.

\begin{algorithm}[tb]
\caption{Rparse: use Rsync to build an LZSS-like parse for a string $S$}
\label{alg:rparse}
\begin{enumerate}
\item build an Rsync dictionary and parse for $S$;
\item append a unique separator character to each block in the dictionary and concatenate the blocks (in the order of their first appearances in $S$) into a string $D$;
\item compute the LZSS parse of $D$;
\item compute the LZSS parse of the parse $P$, treating each block as a meta-character;
\item map $D$'s and $P$'s parses onto $S$:
\label{item:map}
\begin{enumerate}
\item discard any separator character $D [j]$ in $D$;
\item turn the first occurrence $D [j]$ of any other character in $D$ into the first occurrence $S [j']$ of that character in $S$;
\item turn each phrase $D [j..j + \ell - 1]$ in block $B$ with source $D [i..i + \ell - 1]$ in block $B'$, into a phrase $S [j'..j' + \ell - 1]$ with source $S [i'..i' + \ell - 1]$, where $S [j']$ and $S [i']$ have the same respective offsets from the beginnings of the first occurrences of $B$ and $B'$ in $S$, as $D [j]$ and $D [i]$ have from the beginnings of $B$ and $B'$ in $D$;
\item discard the first occurrence $P [j]$ of each block in $P$;
\item turn each phrase $P [j..j + \ell - 1]$ with source $P [i..i + \ell - 1]$, into a phrase $S [j'..j' + \ell' - 1]$ with source $S [i'..i' + \ell' - 1]$, where $S [j']$ and $S [i']$ are the first characters in the $j$th and $i$th blocks, respectively, and $\ell'$ is the total length of the $j$th through $(j + \ell - 1)$st blocks (and thus also the total length of the $i$th through $(i + \ell - 1)$st blocks).
\end{enumerate}
\end{enumerate}
\end{algorithm}

\section{Analysis}
\label{sec:analysis}

The main advantage of using Rsync parsing to preprocess $S$ is that Rsync parsing is quite easy to parallelize, apply over streamed data, or apply in external memory.  The resulting dictionary and parse may be significantly smaller than $S$, making it easier to apply grammar-based compression.  In the full version of this paper we will analyze how much time and workspace Algorithms~\ref{alg:rpair} and~\ref{alg:rparse} use in terms of the total size of the dictionary and parse, but for now we are mainly concerned with the quality of the compression.

Let $b$ be the number of distinct blocks in the Rsync parse of $S$, and let $z$ be the number of phrases in the LZ77 parse of $S$.  The first block is obviously the first occurrence of that substring and if $S [i..j]$ is the first occurrence of another block, then $S [i - w..j]$ (i.e., the block extended backward to include the previous triggering substring) is the first occurrence of that substring.  Since the first occurrence of any non-empty substring overlaps or ends at a phrase boundary in the LZ77 parse, we can charge $S [i..j]$ to such a boundary in $S [i - w..j]$.  Since blocks have length at least $w$ and overlap by only $w$ characters when extended backwards, each boundary has the first occurrences of at most two blocks charged to it, so $b = O (z)$.

In Step~\ref{item:map} of Algorithm~\ref{alg:rparse}, we discard $O (b)$ of the phrases of the phrases in the LZSS parses of $D$ and $P$ when mapping to the phrases in the LZSS-like parse of $S$.  Therefore, by showing that the number of phrases in the LZSS-like parse of $S$ is $O (z)$, we show that the total number of phrases in the LZSS parses of $D$ and $P$ is also $O (z + b) = O (z)$, so the total number of phrases in their LZ77 parses is $O (z)$ as well.

Due to space constraints, the proofs of the results below are in Appendix~\ref{sec:proofs}.

\vspace*{-2mm}
\begin{lemma}
\label{lem:phrases}
If the $t$-th phrase in the LZSS parse of $S$ is $S [j..j + \ell - 1]$ then the $5 t$-th phrase resulting from Algorithm~\ref{alg:rparse}, if it 
exists, ends at or after $S [j + \ell - 1]$.
\end{lemma}

\vspace*{-2mm}
We note that we can quite easily can reduce the five in Lemma~\ref{lem:phrases}, at the cost of complicating our algorithm slightly, but this is not a priority for us right now and we leave it for the full version of this paper.

\vspace*{-2mm}
\begin{corollary}
\label{cor:phrases}
Algorithm~\ref{alg:rparse} yields an LZSS-like parse of $S$ with at most five times as many phrases as its LZSS parse.
\end{corollary}

\vspace*{-3mm}
\begin{theorem}
\label{thm:phrases}
Algorithm~\ref{alg:rparse} yields an LZSS-like parse of $S$ with $O (z)$ phrases.
\end{theorem}

\vspace*{-3mm}
\begin{corollary}
\label{cor:DandP}
The LZ77 parses of $D$ and $P$ have $O (z)$ phrases.
\end{corollary}

\vspace*{-2mm}
Let $A$ be any algorithm that builds an SLP at most an $\alpha$-factor larger than the LZ77 parse of its input.  For example, with Rytter's construction~\cite{Ryt03} we have $\alpha = O (\log (|S| / z))$.

By Corollary~\ref{cor:DandP}, applying $A$ to $D$ --- Step~\ref{item:SLP_D} in Algorithm~\ref{alg:rpair} --- yields an SLP for $D$ with $O (\alpha z)$ rules.  As explained in Section~\ref{sec:algorithms}, Steps~\ref{item:deletion} to~\ref{item:Xi} then increase the number of rules by at most one while modifying the SLP such that, for each block in the dictionary, there is a non-terminal whose expansion is that block.

Similarly, applying $A$ to $P$ --- Step~\ref{item:SLP_P} --- yields an SLP for $P$ with $O (z)$ rules.  Replacing the terminals in the SLP by the non-terminals generating the blocks and then combining the two SLPs --- Steps~\ref{item:replace} and~\ref{item:combine} --- yields an SLP for $S$ with $O (\alpha z)$ rules.  This gives us our main result of this section:

\begin{theorem}
\label{thm:main}
Using $A$ in Steps~\ref{item:SLP_D} and~\ref{item:SLP_P} of Algorithm~\ref{alg:rpair} yields an SLP for $S$ with $O (\alpha z)$ rules.
\end{theorem}

\section{Experiments}
\label{sec:experiments}

We use two genome collections in our experiments: \textsf{c$N$} consists of $N$ concatenated copies of the human chromosome chr19, of about 59MB each; \textsf{s$N$} consists of $N$ concatenated copies of salmonella genomes, of widely different sizes. 

We compare two grammar compressors: \textsf{RePair} \cite{LM00} produces the best known compression ratios but uses a lot of main memory space, whereas \textsf{SOLCA} \cite{TakabatakeIS2017} aims at optimizing main memory usage. Their versions combined with prefix-free parsing are \textsf{BigRepair} and \textsf{BigSOLCA}. \textsf{RePair} could be run only on the smaller collections. Appendix~\ref{sec:setup} gives more details on the experimental setup.

\begin{table}[t]
\begin{center}
\begin{tabular}{l|r||r|r|r|r|r|r||r|r|r|r|r|r}
File & Size & \multicolumn{3}{c|}{\textsf{RePair}} & \multicolumn{3}{c||}{\textsf{BigRePair}} & \multicolumn{3}{c|}{\textsf{SOLCA}} & \multicolumn{3}{c}{\textsf{BigSOLCA}} \\
\hline
     &   & Ratio & Time & Spc\, &  Ratio & Time & Spc\,\, & Ratio & Time & Spc\,\, & Ratio & Time & Spc\,\, \\
\hline
\textsf{c50} & 2.75 & 0.80\% & 1832 & 3842 & 0.91\% & 66.40 & 454.7 & 
1.35\% & 244.1 & 107.4 & 1.54\% & 103.6 & 182.9 \\
\textsf{c100} & 5.51 & 0.30\% & 7311 & 3155 & 0.48\% & 62.13 & 246.4 & 
0.77\% & 236.4 & 53.67 & 0.86\% & 94.03 & 128.8 \\
\textsf{c250} & 13.8 & & & & 0.23\% & 59.95 & 119.8 & 
0.40\% & 239.0 & 29.78 & 0.44\% & 86.39 & 95.00 \\
\textsf{c500} & 27.5 & & & & 0.14\% & 59.97 & 118.0 &
0.28\% & 237.4 & 17.05 & 0.30\% & 85.12 & 84.72 \\ 
\textsf{c1000} & 55.1 & & & & 0.10\% & 60.95 & 117.3 & 
0.22\% & 237.3 & 13.56 & 0.23\% & 86.13 & 78.82 \\
\hline
\textsf{s815} & 3.75 & 1.72\% & 8478 & 3726 & 1.93\% & 90.87 & 2254 & 
3.01\% & 317.7 & 161.0 & 3.50\% & 143.2 & 291.0 \\
\textsf{s2073} & 9.72 & & & & 2.01\% & 95.86 & 1055 &
3.01\% & 370.9 & 153.1 & 3.53\% & 157.3 & 285.5 \\
\textsf{s4570} & 22.0 & & & & 2.61\% & 244.1 & 534.2 & 
3.57\% & 480.6 & 154.4 & 4.24\% & 185.6 & 334.7 \\
\textsf{s11264} & 53.1 & & & & 1.51\% & 2605 & 294.2 &
2.20\% & 620.2 & 92.60 & 2.61\% & 157.3 & 206.4 \\
\end{tabular}
\end{center}

\caption{Performance of the compressors. File sizes are expressed in GB, compression ratios in percentage of compressed file over uncompressed file, compression times in seconds per input GB, and compression main memory usage in MBs per input GB.}
\label{tab:results}
\end{table}

Table~\ref{tab:results} shows the results in terms of compression ratio, time, and space in RAM. On the more repetitive chr19 genomes, \textsf{BigRePair} is clearly the best choice for large files. It loses to \textsf{RePair} in compression ratio, but \textsf{RePair} took 11 hours just to process 5.5GB, so it is not a choice for larger files. Instead, \textsf{BigRepair} processed 55GB in less than an hour and 650MB of RAM. Similarly, \textsf{SOLCA} obtains better compression but more compression time than \textsf{BigSOLCA}, though the latter uses more space. The comparison between the two compressors shows that  \textsf{BigRepair} performs better than both \textsf{SOLCA} and \textsf{BigSOLCA} in both compression ratio (reaching nearly half the compressed size of \textsf{SOLCA} on the largest files) and time ($2/3$ of the time of \textsf{BigSOLCA}). Still \textsf{SOLCA} uses much less space: it compresses 55GB in 3.6 hours, but using less than 75MB.

The results start similarly on the less compressible salmonella collection, but it reaches an important turning point. The time of \textsf{BigRePair} on chr19 was stable around 1GB per minute, but on salmonella it is not: When moving from 10GB to 20GB of input data, the time per processed GB of \textsf{BigRePair} jumps by a factor of 2.5, and when moving from 20GB to 50GB it jumps by more than 10. To process the largest 53GB file, \textsf{BigRePair} requires more than 38 hours and over 15 GB of RAM. \textsf{SOLCA}, instead, handles this file in nearly 9 hours and less than 5 GB, and \textsf{BigSOLCA} in less than 2.5 hours and 11 GB, being the fastest. What happens is that, being less compressible, the output of the prefix-free parse is still too large for \textsf{RePair}, and thus it slows down drastically as soon as it cannot fit its structures in main memory. The much lower memory footprint of \textsf{SOLCA}, instead, pays off on these large and less compressible files, though its compression ratio is worse than that of \textsf{BigRePair}.

\appendix

\section{Omitted Proofs} \label{sec:proofs}

\subsection{Proof of Lemma~\ref{lem:phrases}}

\begin{proof}
Our claim is trivially true for $t = 1$, since the first phrases in both parses are the single character $S [1]$, so let $t$ be greater than 1 and assume our claim is true for $t - 1$, meaning the $5 (t - 1)$st phrase in our parse ends at $S [k - 1]$ with $k \geq j$.  If $k \geq j + \ell$ then our claim is also trivially true for $t$, so assume $j \leq k < j + \ell$.  We must show that our parse divides $S [k..j + \ell - 1]$ into at most five phrases, in order to prove our claim for $t$.

First suppose that $S [k..j + \ell - 1]$ does not completely contain a triggering substring, so it overlaps at most two blocks.  (It can overlap two blocks without containing a triggering substring if and only if a prefix of length less than $w$ lies in one block and the rest lies in the next block.)  Let $S [i..i + \ell - 1]$ be $S [j..j + \ell - 1]$'s source and let $k' = i + k - j$, so in the LZSS parse $S [k..j + \ell - 1]$ is copied from $S [k'..i + \ell - 1]$.  Since $S [k'..i + \ell - 1]$ does not completely contain a triggering substring either, it too overlaps at most two blocks.

Without loss of generality (since the other cases are easier), assume $S [k..j + \ell - 1]$ and $S [k'..i + \ell - 1]$ each overlap two blocks and they are split differently:  $S [k..k + d - 1]$ lies in one block and $S [k + d..j + \ell - 1]$ lies in the next, and $S [k'..k' + d' - 1]$ lies in one block and $S [k' + 1..i + \ell - 1]$ in the next, with $d \neq d'$.  Assume also that $d < d'$, since the other case is symmetric.  Since $S [k..k + d - 1]$ is completely contained in a block and occurs earlier completely contained in a block, as $S [k'..k' + d - 1]$, our parse does not divide it.  Similarly, since $S [k + d..k + d' - 1]$ and $S [k + d'..j + \ell - 1]$ are each completely contained in a block and occur earlier each completely contained in a block, as $S [k' + d..k' + d' - 1]$ and $S [k' + d..i + \ell - 1]$, respectively, our parse does not divide them.  Therefore, our parse divides $S [k..j + \ell - 1]$ into at most three phrases.

Now suppose the first and last triggering substrings completely contained in $S [k..j + \ell - 1]$ are $S [x..x + w - 1]$ and $S [y..y + w - 1]$ (possibly with $x = y$).  By the arguments above, our parse divides $S [k..x + w - 1]$ into at most three phrases.  Since $S [x + w..y + w - 1]$ is a sequence of complete blocks that have occurred earlier (in $S [k'..i + \ell - 1]$), our parse does not divide it unless $S [k..x + w - 1]$ is a complete block that has occurred before as a complete block, in which case it may divide $S [k..y + w - 1]$ once between $S [x + w]$ and $S [y + w - 1]$.  Since $S [y + w..j + \ell - 1]$ is completely contained in a block and occurs earlier completely contained in a block (in $S [k'..i + \ell - 1]$), our parse does not divide it.  Therefore, our parse divides $S [k..j + \ell - 1]$ into at most five phrases. \qed
\end{proof}

\subsection{Proof of Corollary~\ref{cor:phrases}}

\begin{proof}
If the LZSS parse has $t$ phrases then the $t$-th phrase ends at $S [n]$ so, by Lemma~\ref{lem:phrases}, Algorithm~\ref{alg:rparse} yields a parse with at most $5 t$ phrases. \qed
\end{proof}

\subsection{Proof of Theorem~\ref{thm:phrases}}

\begin{proof}
It is well known that the LZSS parse of $S$ has at most twice as many phrases as the its LZ77 parse (since dividing each LZ77 phrase into a prefix with an earlier occurrence and a mismatch character yields an LZSS-like parse with at most twice as many phrases, and the LZSS parse has the fewest phrases of any LZSS-like parse).  Therefore, by Corollary~\ref{cor:phrases}, Algorithm~\ref{alg:rparse} yields a parse with at most $O (z)$ phrases. \qed
\end{proof}

\subsection{Proof of Corollary~\ref{cor:DandP}}

\begin{proof}
Immediate, from Theorem~\ref{thm:phrases}, the fact that the LZ77 parse is no larger than the LZSS parse, and inspection of Algorithm~\ref{alg:rpair}. \qed
\end{proof}

\section{Experimental setup} \label{sec:setup}

Our experiments ran on a Intel(R) I7-4770 @ 3.40 GHz machine with 32 GB memory.

The chr19 collection was downloaded from the 1000 Genomes Project. Each chr19 sequence was derived by using the {\sf bcftools} consensus tool to combine the haplotype-specific (maternal or paternal) variant calls for an individual with the chr19 sequence in the GRCH37 human reference.
The salmonella genomes were downloaded from NCBI (BioProject
PRJNA183844) and preprocessed by assembling each individual sample with IDBA-UD \cite{PLYC12} setting kMaxShortSequence to 1024 per public advice from the author to accommodate the longer paired end reads that modern sequencers produce.
More details of the collections are available in previous work \cite[Sec.~4]{BoucherGKM18}.

For \textsf{RePair} we use Navarro's implementation for large files, at \texttt{http://www. dcc.uchile.cl/gnavarro/software/repair.tgz}, letting it use 10GB of main memory, whereas the implementation of \textsf{SOLCA} is at \texttt{https://github.com/ tkbtkysms/solca}. 
To measure their compression ratios in a uniform way, we consider the following encodings of their output: if \textsf{RePair} produces $r$ (binary) rules and an initial rule of length $c$, we account $2r$ bits to encode the topology of the pruned parse tree (where the nonterminal ids become the preorder of their internal node in this tree) and $(r+c) \lceil \log_2 r \rceil$ bits to encode the leaves of the tree and the initial rule. \textsf{SOLCA} is similar, with $c=1$.

Our code is available at \texttt{https://gitlab.com/manzai/bigrepair}.

\end{document}